\documentclass[12pt]{article}

\usepackage{graphicx}
\usepackage{epsfig}
\usepackage{amsfonts}
\usepackage{amssymb}
\usepackage{color}
\usepackage{cite}
\usepackage{amsmath,amssymb}

\usepackage{hyperref} 

\newcommand{\ee}{\end{equation}}
\newcommand{\eea}{\end{eqnarray}}
\newcommand{\be}{\begin{equation}}
\newcommand{\bea}{\begin{eqnarray}}

\begin{document}

\title{  
 {Einstein-(complex)-Maxwell static boson stars in AdS}
}
 \vspace{1.5truecm}
 %\date{\ch{May 2024}}
 \date{}
%\pacs{04.20.Jb, 04.40.Nr}

\author{
{\large Carlos Herdeiro}$^{1}$,
{\large Hyat Huang}$^{2,3}$,
{\large Jutta Kunz}$^{2}$
and
{\large Eugen Radu}$^{1}$
\\
$^{1 }${\small Departamento de Matem\'atica da Universidade de Aveiro and } 
\\ {\small  Centre for Research and Development  in Mathematics and Applications (CIDMA),} 
\\ {\small    Campus de Santiago, 3810-183 Aveiro, Portugal}
\\ 
$^{2}${\small Institute of Physics, Carl von Ossietzky University Oldenburg,
Germany Oldenburg D-26111, Germany}
\\ 
$^{3}${\small 
College of Physics and Communication Electronics, Jiangxi Normal University,
Nanchang 330022, China}
}

\maketitle

\begin{abstract}
We consider a model with two real  Maxwell fields  (or equivalently, a complex Maxwell field) minimally 
coupled to Einstein’s gravity with a negative cosmological constant in four spacetime dimensions.
Assuming a specific harmonic dependence of the vector fields,
we show the existence of
 asymptotically anti-de Sitter (AdS) self-gravitating boson-star--like solitonic solutions,
which are static and axially symmetric.
Analytical solutions are found in the test-field limit, 
where the Maxwell equations are solved  on  a fixed AdS background.
The fully nonlinear solutions are constructed numerically.
\end{abstract}
 
 \vfill {\footnotesize 
~\\
herdeiro@ua.pt\\
hyat@mail.bnu.edu.cn\\
jutta.kunz@uni-oldenburg.de\\
eugen.radu@ua.pt
}\ \ \ \
 
%\tableofcontents

%%%%%%%%%%%%%%%%%%%%%%%%%%%%%%%%%%%%%%%%%%%%%%%%%%%%%%%%%%%%%%%%%%%%%%%%%%%%%%
\section{Introduction}
%%%%%%%%%%%%%%%%%%%%%%%%%%%%%%%%%%%%%%%%%%%%%%%%%%%%%%%%%%%%%%%%%%%%%%%%%%%%%%
 
Spacetimes which asymptotically
behave as anti-de Sitter ($AdS$) have attracted significant
interest since the proposal of the $AdS$/conformal field theory 
duality \cite{Maldacena:1997re}, 
which asserts that a theory of quantum gravity in $D-$dimensions has
 a dual formulation in terms of a non-gravitational theory in $(D-1)$-dimensions.
Moreover, and independently of this, being a maximally
symmetric geometry, the $AdS$ geometry provides a useful
 background to investigate questions of principle related to the
 behaviour of classical or quantum fields on a non-trivial background, as well as, for instance,
 the status of the black hole no hair conjecture~\cite{Ruffini:1971bza} 
when relaxing the assumption of asymptotic flatness. 

 The peculiar AdS asymptotics indeed allow new possibilities.
 For example, as shown in 
Refs.
\cite{Herdeiro:2015vaa,Costa:2015gol,Herdeiro:2016xnp,Herdeiro:2016plq},
electrostatics on $D=4$ global $AdS$ has some very different features from
standard electrostatics on Minkowski spacetime. 
 A striking illustration  is that all multipole moments 
(except for the monopole) possess everywhere regular, finite energy configurations, defined by their multipole structure at the AdS boundary.  
Their non-linear backreacting versions 
yield Einstein-Maxwell-$AdS$ solitons, 
which inherit the spatial symmetries of the boundary data. 
Moreover, introducing a horizon yields a static black hole \cite{Costa:2015gol},
which, for appropriate boundary multipoles,
has no continuous spatial symmetries \cite{Herdeiro:2016plq}.

\medskip

The study in Refs. 
\cite{Herdeiro:2015vaa}-\cite{Herdeiro:2016plq}
 was restricted to the case of a single
Maxwell field, with no time dependence.
In this work we enlarge the framework therein by considering a model with
$two$ Maxwell fields,  
which
are endowed with a harmonic time dependence.
However, the time dependence
 disappears at the level of the total energy momentum tensor,
such that the considered configurations
are static.
As such, the solutions resemble 
 the well-known (scalar or vector) boson stars (BSs)  
 \cite{Liebling:2012fv}. 
This can be understood by noticing that the 
two Maxwell fields 
  can be taken as a (single) complex vector field,
 the considered model  corresponding to the massless limit 
	of the Einstein-Proca-$AdS$ theory.  
The spherically symmetric asymptotically  $AdS$
 solutions of a model with
a complex massive spin-1 field  minimally coupled to
Einstein's gravity with negative cosmological constant were studied
in Ref. \cite{Duarte:2016lig}.
These $AdS$ Proca stars 
 share very similar properties to the spin zero $AdS$ boson stars
\cite{Astefanesei:2003qy,Buchel:2013uba}, 
$e.g.$ the dependence of their global charges
on the field frequency.

\medskip

In this work we construct and 
discuss the basic properties
of the simplest BS-like solutions  
 of the $D=4$
 Einstein-(complex)-Maxwell (EcM)  model with a negative cosmological constant.
Our results show that,
despite the harmonic time dependence of the vector potential  in both cases,
the  configurations here
 exhibit a rather different picture as compared to that found
in Ref. \cite{Duarte:2016lig} for $AdS$ massive vector fields.
For example, no  everywhere regular spherically symmetric solitons exist in the
 EcM  case, while the solutions possess a non-trivial zero-frequency limit. 

\medskip
This paper is organized as follows. 
In Section~\ref{sec_2} we present the  EcM model.
In Section~\ref{sec_3} we discuss the (everywhere regular) solitonic solutions on a fixed 
$AdS$ background (dubbed \textit{clouds}) for a specific ansatz with a magnetic potential possessing
a harmonic time dependence.
 Then, in Section~\ref{sec_4}  the backreaction on $AdS$ of
a particular class of
 clouds is considered.
We construct, both perturbatively (analytically) and non-perturbatively (numerically) 
the corresponding  EcM-$AdS$ solitons.  
In Section~\ref{sec_5} we draw our conclusions and further remarks. 

%%%%%%%%%%%%%%%%%%%%%%%%%%%%%%%%%%%%%%%%%%%%%%%%%%%%%%%%%%%%%%%%%%%%%%%%%%%%%%
\section{The EcM model with negative cosmological constant}
\label{sec_2}
%%%%%%%%%%%%%%%%%%%%%%%%%%%%%%%%%%%%%%%%%%%%%%%%%%%%%%%%%%%%%%%%%%%%%%%%%%%%%%
 
We consider the (four dimensional)
Einstein gravity with negative cosmological constant $\Lambda=-3/L^2$
coupled with two real Maxwell fields  
% complex massless vector field
\begin{eqnarray}
\label{action}
\mathcal{S} = \frac{1}{4 \pi G} \int d^4 x\sqrt{-g}
\left[
\frac{1}{4}
\left(R+\frac{6}{L^2}\right)
-\frac{1}{4}
F_{\mu \nu}^{(1)}F^{(1)\mu\nu}
-\frac{1}{4}
F_{\mu \nu}^{(2)}F^{(2)\mu\nu}
%F_{\mu \nu}F^{*\mu\nu}
\right] ,
\end{eqnarray}
where $F^{(a)}=dA^{(a)}$, $a=1,2$.
 
After defining 
\begin{eqnarray}
\mathcal{F}=F^{(1)}+i F^{(2)},
\end{eqnarray}
one can write an equivalent form of
(\ref{action}),
with a matter content 
corresponding to a
$complex$ massless vector field
$\mathcal{F}$:
\begin{eqnarray}
\label{actionN}
\mathcal{S} = \frac{1}{4 \pi G} \int d^4 x\sqrt{-g}
\left[
\frac{1}{4}
\left(R+\frac{6}{L^2}\right)
-\frac{1}{4} 
 \mathcal{F}_{\mu \nu}\bar{ \mathcal{F}^{\mu\nu}}
\right]
,
\end{eqnarray}
with $\mathcal{F}=d\mathcal{A}$
and $\mathcal{A}=A^{(1)}+i A^{(2)}$,
while an overbar  denotes the complex conjugate.

For this formulation,
the Einstein equations are
\begin{equation}
\label{Einstein-eqs}
R_{\alpha \beta}
-\frac{1}{2}R g_{\alpha \beta} 
-\frac{3}{L^2} g_{\alpha \beta}=2~ T_{\alpha \beta}~,
\end{equation}
with the energy-momentum tensor
\begin{eqnarray}
\label{Tik}
T_{\alpha\beta}=\frac{1}{2}
( \mathcal{F}_{\alpha \sigma }\bar{\mathcal{F}}_{\beta \gamma}
+\bar{\mathcal{F}}_{\alpha \sigma } \mathcal{F}_{\beta \gamma}
)g^{\sigma \gamma}
-\frac{1}{4}g_{\alpha\beta}\mathcal{F}_{\sigma\tau}\bar{\mathcal{F}}^{\sigma\tau}~.
\end{eqnarray}
Also,
the (complex) four-potential $\mathcal{A}$ satisfies the  equations
\begin{eqnarray}
\label{F-eqs}
\nabla_\mu \mathcal{F}^{\mu \nu}=0.
\end{eqnarray}

Let us remark that
the action (\ref{actionN})
can be taken as the zero  field  mass limit of the Einstein-complex-Proca-AdS 
system.
As with that case, the model
possess a 
$global$ 
$U(1)$ invariance  
 under the transformation $\mathcal{A}_\mu\rightarrow e^{i\alpha}\mathcal{A}_\mu$ 
(with $\alpha$ a constant), 
which
implies the existence of a conserved 4-current, 
\begin{equation}
\label{current}
j^\alpha=\frac{i}{2}\left[\bar{\mathcal{F}}^{\alpha \beta}\mathcal{A}_\beta-\mathcal{F}^{\alpha\beta}\bar{\mathcal{A}}_\beta\right] ,
\end{equation}
with $\nabla_\alpha j^\alpha=0$. 
Consequently, 
even for a massless field $\mathcal{A}$
there exists a Noether charge $Q$ (the particle number), 
obtained integrating the temporal component of the 4-current on a space-like slice $\Sigma$:
\begin{equation}
Q=\int_\Sigma d^3x ~\sqrt{-g}~j^t \ .
\label{q}
\end{equation}
 However,  
different from the Proca case,
% the local $U(1)$ invariance of each U(1) field 
%s still preserved
 the model (\ref{actionN}) is invariant also 
under the $local$ transformation
$\mathcal{A}_\mu\rightarrow  \mathcal{A}_\mu+\partial_\mu \beta(x^\gamma)$,
with $\beta$ a complex function of spacetime coordinates.

As discussed in Ref. \cite{Duarte:2016lig},
the Einstein-Proca model with negative cosmological constant 
possesses 
(spherically symmetric) 
solitonic solutions.  
This brings up the question if 
such solutions exist as well when taking the limit of a massless  complex vector field, keeping the harmonic time dependence.
However, once can prove that 
this is not the case when considering
 spherically symmetric configurations\footnote{
This can be proven by considering the usual ansatz for
 spherically symmetric configurations \cite{Brito:2015pxa},
with
a vector potential
$
\mathcal{A}=\left[f(r)dt+ig(r)dr\right] e^{-i \omega t},
$
and a line element
$ds^2= g_{tt}(r)dt^2+ g_{rr}(r)dr^2+r^2(d \theta^2+\sin^2\theta d\varphi^2)$,
with
$g_{rr}(r),g_{tt}(r);f(r),g(r)$ real functions of the radial coordinate $r$
and $\omega$ the real frequency parameter.
As such, the only non-zero components of the field strength tensor are
$\mathcal{F}_{rt}=-\mathcal{F}_{rt}= (f'(r)-\omega g(r)) e^{-i \omega t}$.
However,  
 they are identically zero, as implied by the  field equation 
$\nabla_\alpha \mathcal{F}^{\alpha r}=0$.};
in particular, the Proca stars in Ref. \cite{Duarte:2016lig} trivialize 
for a massless vector field.
However, this does not exclude the existence of static, everywhere regular non-symmetric configurations, 
as proven in the next Sections. 

%%%%%%%%%%%%%%%%%%%%%%%%%%%%%%%%%%%%%%%%%%%%%%%%%%%%%%%%%%%%%%%%%%%%%%%%%%%%%%
\section{The probe limit:  a massless complex  magnetic vector field  on  global $AdS_4$}
\label{sec_3}
%%%%%%%%%%%%%%%%%%%%%%%%%%%%%%%%%%%%%%%%%%%%%%%%%%%%%%%%%%%%%%%%%%%%%%%%%%%%%%
 
As a first step towards solving the full set of  EcM  equations, 
we shall consider a simplified version of the model 
where we
ignore the Maxwell field  backreaction on the spacetime geometry.
That is, one solves the vector field equations (\ref{F-eqs}) on
 a fixed {\it global} $AdS_4$ background with a line element
\begin{eqnarray}
\label{AdS}
ds^2=-N(r)dt^2+\frac{dr^2}{N(r)}+r^2(d \theta^2+\sin^2\theta d\varphi^2)\ ,~~
{\rm and}~~N(r)=1 +\frac{r^2}{L^2}\ ,
\end{eqnarray} 
%with the cosmological constant $\Lambda=-3/L^2$.
where $r$ and $t$
are the radial and time coordinates, while 
$\theta,\varphi$
are the usual angular variables on $S^2$).  
The aim is to 
look  for {\it vector clouds},
$i.e.$ configurations which are regular everywhere and display
no time-dependence at the level of  the  total energy-momentum tensor (\ref{Tik}).
%
  
%Therefore one  has to consider less symmetric  configurations.
 
Restricting to  axial symmetry, 
the simplest complex vector ansatz has a purely magnetic potential, 
with\footnote{The same ansatz in terms of the real potentials $A^{(a)}$
reads 
$A^{(1)}=\Phi(r,\theta) \cos \omega t$,
$A^{(2)}=\Phi(r,\theta) \sin \omega t$,
which corresponds two waves with a $\pi/2$ phase difference.
}
\begin{eqnarray}
\label{gauge-ansatz}
{\cal A}={\cal A}_\mu dx^\mu= \Phi(r,\theta) e^{-i \omega t}   d \varphi,
\end{eqnarray}  
with $\omega>0$ the frequency and $\Phi$ the (real) field amplitude.
This results in the following  non-vanishing
components of the field strength tensor
$
\mathcal{F}_{r \varphi}=-\mathcal{F}_{  \varphi r}= \Phi_{,r}  e^{-i \omega t} 
$,
$
\mathcal{F}_{\theta \varphi}= -\mathcal{F}_{\varphi \theta }=\Phi_{,\theta}  e^{-i \omega t},
$
and
$
\mathcal{F}_{ \varphi t}= -\mathcal{F}_{ t \varphi}= i \omega \Phi  e^{-i \omega t}.
$

As with the usual Proca stars,
 the energy-momentum tensor possesses no time dependence. 
The energy    
and the Noether charge densities
  of these configurations are
\begin{eqnarray}
\rho=-T_t^t=
\frac{1}{2r^2\sin^2\theta}
\left(
N\Phi_{,r}^2+\frac{\Phi_{,\theta}^2}{r^2}
+\frac{\omega^2 \Phi^2}{N}
\right) ,~~~j^t=\frac{ \omega \Phi^2}{ r^2 N\sin^2\theta}~.
\end{eqnarray}
From (\ref{F-eqs}),
the equation for the magnetic potential amplitude
$\Phi$ reads:
\begin{eqnarray}
\label{gauge-eq}
 \Phi_{,rr}
+\frac{ \Phi_{,\theta \theta}}{r^2 N s}
+\frac{ N' \Phi_{,r}}{r^2 N}
-\frac{ \cot \theta \Phi_{,  \theta}}{r^2 N} 
+\frac{  \omega^2 \Phi }{  N^2}=0~.
\end{eqnarray}  
We
assume separation of variables,
such that $\Phi$  can be written as a sum of modes,
\begin{eqnarray}
\label{phi-gen}
\Phi(r,\theta )= \sum_{k \geq 1} s_k R_k(r) U_{k}( \theta)\ ,~~{\rm where}~~
  U_{k}( \theta)=\sin \theta \frac{d  \mathcal{P}_k(\cos \theta)}{d\theta} ~,
\end{eqnarray}
 $\mathcal{P}_k$ being a Legendre polynomial of degree $k$.
It follows that radial amplitude  $R_k(r)$ of a mode $k$ 
solves the equation
\begin{eqnarray}
\label{eq}
 N(N R_k')'+ \left( \omega^2-\frac{k (k+1)N}{r^2}  \right)R_k=0~.
\end{eqnarray}
The solution
of the above equation 
which is regular everywhere (in particular at $r=0$)
reads:
\begin{eqnarray}
\label{sol-gen}
R_k(r)=
c_0 \left(\frac{r}{L} \right)^{1+k} (1+\frac{r^2}{L^2})^{\frac{L \omega}{2}}~
{}_2F_1 
\left (
\frac{1}{2}(1+k+L \omega),
 \frac{1}{2}(2+k+L \omega),\frac{3}{2}+ k,- \frac{r^2}{L^2}
\right ),
\end{eqnarray} 
with  ${}_2F_1$  the hypergeometric function, 
where $c_0$
is an arbitrary (nonzero) constant. 
Note that 
the (magnetic) Maxwell clouds  reported in
a larger context in 
Ref. \cite{Herdeiro:2016xnp}
are recovered
in the $\omega \to 0$ limit.
As with that case, the radial amplitude
{\it does not}  generically  vanishes as $r\to \infty$,
approaching a constant value which in our case is a function of $\omega$ and $L$.
A natural choice for $c_0$
is to impose
$R_k(r) \to 1$ as
$r \to \infty$,
which results in
\begin{eqnarray}
c_0 =\frac{\Gamma(\frac{1}{2}(2+k-L \omega))\Gamma(\frac{1}{2}(2+k+L \omega))}
{\sqrt{\pi} \Gamma(\frac{3}{2}+k) }.
\end{eqnarray} 
%
%The explicit form of the radial amplitude
%  for the first three values of $k$ reads
 For completeness, we display the 
 simplified form of the first two radial 
 amplitudes,
\begin{eqnarray}
&&
\nonumber
R_1(r)= \frac{1}{\cos (\frac{\pi L \omega }{2})} 
\bigg(
\cos \left[L \omega \arctan \left(\frac{r}{L}\right)\right]
-\frac{1}{ \omega r}  \sin \left[L \omega \arctan \left(\frac{r}{L}\right)\right]
\bigg),
\\
&&
\label{sols} 
R_2(r)=\frac{1}{(1-L^2 \omega^2)^2 \sin (\frac{\pi L \omega }{2})}
\sqrt{
 (1-L^2 \omega^2)^2
+ \frac{3L^2(2+L^2 \omega^2)}{r^2}
+\frac{9L^4}{r^4}
} 
\\
\nonumber
&&
{~~~~~~~~~~~}
\times 
\sin \bigg(
L \omega \arctan \left(\frac{r}{L}\right) 
+\arctan \bigg[ \frac{ r \omega}{(L^2 \omega^2-1)\frac{r^2}{3L^2}-1 }\bigg]
\bigg)~.
%\\
%&&
%\nonumber
%R_3(r)=\frac{L^2}{r^2 (L^2\omega^2-4)\cos(\frac{\pi L \omega}{2})}
%\bigg( 
%(r^2\omega^2-\frac{4r^2}{L^2}-15 )\cos(L\omega\arctan(\frac{r}{L}))
%\\
%\nonumber
%&&
%{~~~~~~~~~~~}
%+(\frac{3r^2}{L^2}-2r^2\omega^2+5 ) \frac{3}{r \omega}\sin(L\omega\arctan(\frac{r}{L}))
%\bigg). 
\end{eqnarray} 
%the $$ expression for higher $k$ becoming increasingly complicated.
%
As $r\to 0$, the generic solutions behave as
\begin{eqnarray}
R_k(r)=u_0^{(k)} \left(\frac{r}{L}\right)^{k+1}+\dots,
{\rm with}~~
u_0^{(k)}= \frac{2\Gamma(\frac{1}{2}(2+k-L \omega))\Gamma(\frac{1}{2}(2+k+L \omega))}
{ \sqrt{\pi}\Gamma(\frac{3}{2}+k) }  
\end{eqnarray} 
while asymptotically
\begin{eqnarray}
\label{as11}
R_k(r)=1-\mathfrak{ m}^{(k)}\frac{L}{r}+\dots,~~
{\rm with}~~
\mathfrak{ m}^{(k)}=\frac{2\Gamma(\frac{1}{2}(2+k-L \omega))\Gamma(\frac{1}{2}(2+k+L \omega))}
{ \Gamma(\frac{1}{2}(1+k-L \omega))\Gamma(\frac{1}{2}(1+k+L \omega)) }  ~.
\end{eqnarray}

The energy density $\rho$ of the solutions is finite everywhere
and strongly localized in a finite region of space, 
with $\rho$ nonzero at $\theta=0,\pi$, and decaying as $1/r^4$ for large $r$.
The  total mass-energy  and Noether charge of solutions are 
 \begin{eqnarray}
M=- 2\pi \int_0^\infty r^2 dr \int_0^\pi d\theta \sin \theta ~ T_t^t,~~
~~
Q_N=   2\pi  \int_0^\infty r^2 dr \int_0^\pi d\theta \sin \theta  ~j^t ~,
\end{eqnarray} 
%%%%%%%%%%%%%%%%%%%%%%%%%%%%%%%%%%%%%%%%%%%%%%%%%%%%%%%%%%%%%%%%%%%%%%%%%%%%%%
 \begin{figure}[h!]
\begin{center}
\includegraphics[width=0.45\textwidth]{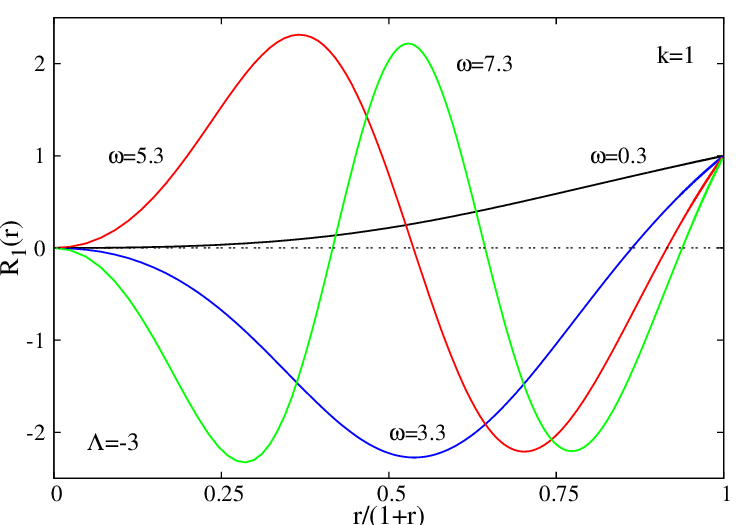}
\includegraphics[width=0.45\textwidth]{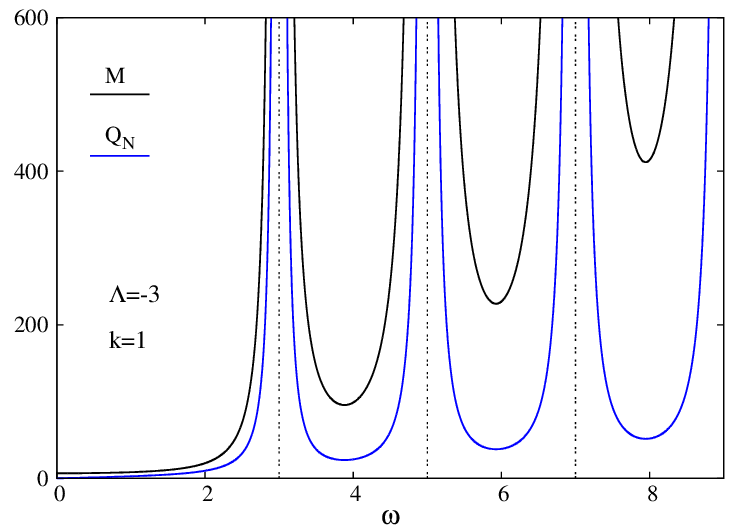}
\caption{
{\it Left:}
The profile of $k=1$ solutions with $n=0,\dots 3$
nodes is shown as a function of compactified radial coordinate.
{\it Right:}
The mass and Noether charge is shown as a function of 
frequency for $k=1$ solutions. 
}
\label{profile}
\end{center}
\end{figure} 
%%%%%%%%%%%%%%%%%%%%%%%%%%%%%%%%%%%%%%%%%%%%%%%%%%%%%%%%%%%%%%%%%%%%%%%%%%%%%%
%
with the following expressions
for the first two values of $k$:
\begin{eqnarray}
&&
\nonumber
M_{(1)}=\frac{2 \pi^2}{3L \cos^2\frac{\pi L\omega}{2}}
\left(
L^2 \omega^2-1+\frac{2}{\pi L\omega}\sin \pi L \omega
\right),
\\
&&
M_{(2)}=\frac{6\pi^2 L\omega^2 }{5 (1-L^2 \omega^2)^2 \sin^2 \frac{\pi L\omega}{2} }  
\left[
(4-L^2\omega^2)(1-L^2\omega^2)-\frac{6L\omega}{\pi}\sin \pi L \omega
\right],
%\\
%&&
%\nonumber
%M_{(3)}=\frac{12\pi^2 }
%{(7L (L^2\omega^2-4)^2)\cos^2(\frac{\pi L\omega}{2})}
%\bigg( L\pi\omega(L^2\omega^2(L^2\omega^2-7)^2-36)+12(6-3L^2\omega^2+L^4\omega^4)\sin\pi L\omega \bigg),
%\bigg( \L^2\omega^2(L^2\omega^2-7)^2-36
%+(6-3L^2\omega^2+L^4\omega^4) \frac{12}{\pi L \omega}\sin\pi L\omega \bigg),
\end{eqnarray} 
and
\begin{eqnarray}
&&
Q_{N(1)}=\frac{2 \pi^2}{3  L \omega \cos^2\frac{\pi L\omega}{2}}
\left(
L^2 \omega^2-1+ \frac{L^2 \omega^2+1}{\pi L \omega}
\sin \pi L \omega
\right),
\\
&&
	\nonumber
Q_{N(2)}=\frac{6 \pi^2 L \omega}{5  (1-L^2 \omega^2)^2 \sin^2 \frac{\pi L\omega}{2} } 
\left[
(4- L^2\omega^2)(1-L^2\omega^2)-\frac{(4+L^2 \omega^2+L^4\omega^4 )}{\pi L \omega}
\sin \pi L \omega
\right]~.
%\\
%&&
%	\nonumber
%
%Q_{N(3)}=\frac{12\pi^2}{7L\omega(L^2\omega^2-4)^2\cos^2(\frac{\pi L \omega}{2})}
%\bigg( L^2\omega^2(L^2\omega^2-7)^2-36  
%\\
%&&
%	\nonumber
%{~~~~~~~~~~~~~~~~~~~~~~~~~~~~~~~~~~~~~~~~~~~~}
%+(L^6\omega^6 -2L^4\omega^4+13L^2\omega^2 +36)
%\frac{\sin \pi L\omega}{\pi L \omega}
%\bigg),
\end{eqnarray} 

One can prove that for a general multipolar distribution,
%for a given mode $k$
%(with asymptotics
%(\ref{as11})),
the mass is the sum of a Noether charge contribution
and a term which can be expressed as a sum of multipolar momenta,
%and a term proportional with the (magnetic) multipole momenta  $\mathfrak{ m}^{(k)}$,
%can be written as
\begin{eqnarray}
%M_{(k)}=\omega Q_{N(k)}+\frac{2k(k+1)\pi}{(2k+1)L}  \mathfrak{ m}^{(k)} ~.
M =\omega Q_{N }+ \sum_{k \geq 1 } s_k^2 \frac{2k(k+1)\pi}{(2k+1)L}\mathfrak{ m}^{(k)} ~.
\end{eqnarray}

\medskip

For the rest of this work, we shall restrict our study 
to the case $k=1$,
which captures already most of the generic features,
and work in units with $L=1$.
In Figure \ref{profile} we display the profile  of 
the radial amplitude $R_1(r)$ for several
value of $\omega$ (left panel), 
together with the mass and Noether charge 
as a function of frequency (right panel).
There one notices the existence of 
a set of {\it critical} frequencies
\begin{eqnarray}
\label{wcrit}
\omega_{c}^{(n)}=\frac{2n+1}{L},
\end{eqnarray}
with $n$ a positive integer.
The radial amplitude $R_1(r)$ 
 possesses $n-1$ nodes
for 
 $\omega_{c}^{(n)}<  \omega <\omega_{c}^{(n+1)}$, $e.g.$ it is nodeless
for $0\leq \omega < 3/L$.

Both the mass and the Noether charge diverge as
$\omega \to \omega_{c}^{(n)}$ - Figure \ref{profile} (right panel)
This divergence  is a consequence of
imposing the radial amplitude to approach 
a nonvanishing value at infinity. 
In fact, the solutions with 
$\omega=\omega_{c}^{(n)}$
form a separate set with some special properties.
In particular, if the radial amplitude
vanishes both at $r=0$ and at infinity, 
the mass and charge become finite with
$R_1^{(n)}(r)$
possessing $n-1$ nodes.
From  
 the general expression (\ref{sol-gen})
(with $c_0=1$) one finds
\begin{eqnarray}
R_1^{(n)}(r)=  \frac{r^2}{L^2}
\frac{ \sum_{m=0}^{n-1}s_{m}^{(n)} (\frac{r}{L})^{2m}}{(1+\frac{r^2}{L^2})^{(2n+1)/2}},
\end{eqnarray} 
where
$s_0^{(n)}=1$,
$s_1^{(1)}=-3/5$,
$s_2^{(1)}=-2$,
$s_2^{(2)}=3/7$, 
$etc$.
The mass of these solutions is
$M_{(1)}^{(n)}=\frac{3\pi^2}{2L}\frac{(n-1)!}{(n+1)!}$,
while
%$M_{(1)}^{(n)}=\frac{2n+1}{L} Q_{N(1)}^{(n)}$.
$Q_{N(1)}^{(n)}=\frac{L}{2n+1} M_{(1)}^{(n)} $.

%%%%%%%%%%%%%%%%%%%%%%%%%%%%%%%%%%%%%%%%%%%%%%%%%%%%%%%%%%%%%%%%%%%%%%%%%%%%%
\section{Including the backreaction}
\label{sec_4}
%%%%%%%%%%%%%%%%%%%%%%%%%%%%%%%%%%%%%%%%%%%%%%%%%%%%%%%%%%%%%%%%%%%%%%%%%%%%%%  
 Similar to the zero frequency limit,
Refs.    
\cite{Herdeiro:2015vaa}-\cite{Herdeiro:2016plq},
the AdS complex vector clouds discussed above 
should
 possess  extensions in the full 
model (\ref{actionN}), 
 $i.e.$ when taking into account the 
backreaction on the spacetime geometry.

%%%%%%%%%%%%%%%%%%%%%%%%%%%%%%%%%%%%%%%%%%%%%%%%%%%%%%%%%%%%%%%%%%%%%%%%%%%%%
\subsection{A perturbative approach}
\label{sec_41}
%%%%%%%%%%%%%%%%%%%%%%%%%%%%%%%%%%%%%%%%%%%%%%%%%%%%%%%%%%%%%%%%%%%%%%%%%%%%%% 

In the perturbative construction  of the axially symmetric EcM solutions, 
it is convenient to consider a generalization of the
pure $AdS$ line element (\ref{AdS}) with three unknown functions $ {\cal U}_i$, 
\be
ds^2=-N(r) {\cal U}_1(r,\theta)dt^2
+\frac{ {\cal U}_2(r,\theta)}{N(r)}dr^2
+{\cal U}_3(r,\theta)r^2 (d \theta^2+\sin^2\theta d\varphi^2),
\ee
while the expression of the vector potential,
 up to order $\mathcal{O}(\epsilon^3)$, is 
 %\ch{[Why not $\epsilon^2?$ Add comment.]}
%
 \begin{eqnarray}
\label{nex2}
{\cal A}={\cal A}_\mu dx^\mu=   \Phi(r,\theta) e^{-i \omega t}   d \varphi
~~{\rm and}~~\Phi(r,\theta) =\epsilon \Phi^{(1)}(r,\theta) +  \epsilon^3 \Phi^{(3)}(r,\theta) +\dots\ , 
\end{eqnarray}
where $\Phi^{(1)}(r,\theta) $ is a linear  vector
on $AdS$ studied in the previous Section 
and $\epsilon$ is an infinitesimally small parameter\footnote{
Formally, one can add a $\epsilon^2 \Phi^{(2)}(r,\theta)$-term
in the expanssipn (\ref{nex2}).
However, 
$\Phi^{(2)}$
solves the same equation as 
$\Phi^{(1)}$ and can be set to zero.
}.
The 
backreaction of the vector field on the geometry is taken into account by
  considering the following ansatz for the metric functions 
\bea 
{\cal U}_i(r,\theta)=1+\epsilon^2 q_{i2}(r,\theta)+\epsilon^4 q_{i4}(r,\theta)+\cdots. 
%~
%U_2=1+\alpha^2 q_{22}(r,\theta)+\alpha^4 q_{24}(r,\theta)+\cdots, 
%~
%U_3=1+\alpha^2 q_{32}(r,\theta)+\alpha^4 q_{34}(r,\theta)+\cdots,
\eea
Then the coupled EcM field equations are solved order by order
in $\epsilon$,
the free constants which enter the solution
being fixed by imposing regularity at 
$r=0$
and standard $AdS$ asymptotics.

\medskip

To illustrate this procedure, 
we consider the backreaction on the geometry of a magnetic dipole cloud,
$i.e.$ the  solution of eq. (\ref{eq}) with $k=1$.
To lowest order, the $\theta$-dependence in ${\cal U}_i$
is factorized by using the 
following consistent ansatz
\bea
 q_{i2}(r,\theta)=a_i(r)+b_i(r)P_2 (\cos \theta ),~~
% q_{22}=a_2(r)+b_2(r)P_2 (\cos \theta)~~
% q_{12}=a_3(r)+b_3(r)P_2 (\cos \theta).
\eea
Also, a residual (metric) gauge freedom is used
 to set 
$a_3=0$.
Then the Einstein equations can be solved to find 
the expression of the remaining functions $a_i(r)$, $b_i(r)$.

Unfortunately, their expression for generic $\omega>0$ is extremely complicated.
However, a simple enough form is found for 
$\omega=0$ 
\cite{Herdeiro:2016xnp,Costa:2015gol},
and for $\omega=(2n+1)/L$, with $n=1,2,\dots$.
For example, the solution with $\omega=3/L$ 
(in which case 
one takes
$\Phi^{(1)}(r,\theta)=\frac{r^2}{L^2}\frac{\sin^2\theta }{N(r)^{3/2}}$)
reads
\bea
&&
\nonumber
a_1(r)=-\frac{L\bigg(3L^2r+5Lr^3+9(L^2+r^2)^2\arctan(r/L)\bigg)}{12r(L^2+r^2)^3},~
b_1(r)=\frac{2r^2(L^4+5L^2r^2-2r^4)}{3(L^2+r^2)^4},~~
\\
&&
a_2(r)=\frac{L\bigg(Lr^3-9L^3r+9(L^2+r^2)^2\arctan(r/L)\bigg)}{12r(L^2+r^2)^3},~
\\
\nonumber
&&
b_2(r)=\frac{2r^2(7L^4-31L^2r^2+4r^4)}{3(L^2+r^2)^4},~~
b_3(r)=\frac{2r^2(7L^2-2r^2)}{3(L^2+r^2)^3}.
\eea
To this order, the total mass of solutions, computed by using
the same approach as with the non-perturbative case below, 
reads 
\begin{eqnarray}
\label{Mpert}
M=\frac{3\pi\epsilon^2 }{16 G L}~.
\end{eqnarray}
In computing the $\epsilon^3$-correction  to the gauge potential induced by the
deformation of the $AdS$ background, one
takes the following decomposition
\begin{eqnarray}
\label{Phipert}
\Phi^{(3)}(r,\theta) =
\phi^{(3,1)}(r) U_1(\theta)+\phi^{(3,2)}(r) U_3(\theta),
\end{eqnarray}
with $U_k(\theta)$ defined by eq. (\ref{phi-gen}). 
The radial functions above are found by solving 
the eq.   (\ref{F-eqs}) with the assumption of regularity,
which results in
\begin{eqnarray}
\label{Phiperts} 
&&
\nonumber
\phi^{(3,1)}(r)=\frac{1}{(L^2+r^2)^{3/2}}
\bigg[
\arctan\left(\frac{r}{L}\right)
\left(
\frac{3L^6+21 L^4 r^2 +63 L^2 r^4+9 r^6} {4L r(L^2+r^2)}
+9 r^2 
\arctan\left(\frac{r}{L}\right)\right) 
\\
&&
{~~~~~~~~~}-\frac{720L^{10}-411 L^8 r^2+1364 L^6 r^4 +17662 L^4 r^6
+9476 L^3 r^8+3525 r^{10}}{960 (L^2+r^2)^4}
\bigg],
\\
&&
\nonumber
\phi^{(3,2)}(r)=-\frac{r^4 (105 L^6-61 L^4 r^2+27 L^2 r^4+r^6}{160L (L^2+r^2)^{11/2}}~.
\end{eqnarray}
%(with $p_1$ a constant of integration).
The computation can be extended to higher order in $\epsilon$;
however, the corresponding expressions become very complicated
and not enlightening, {\it per se}.
 
%%%%%%%%%%%%%%%%%%%%%%%%%%%%%%%%%%%%%%%%%%%%%%%%%%%%%%%%%%%%%%%%%%%%%%%%%%%%%
\subsection{Nonperturbative solutions}
\label{sec_42}
%%%%%%%%%%%%%%%%%%%%%%%%%%%%%%%%%%%%%%%%%%%%%%%%%%%%%%%%%%%%%%%%%%%%%%%%%%%%%% 
Fully non-linear $AdS$-solitons are obtained by directly solving numerically 
the  equations 
(\ref{Einstein-eqs}),
(\ref{F-eqs}).
In what follows, we shall employ
the Einstein--De Turck (EDT) approach, proposed in~\cite{Headrick:2009pv,Adam:2011dn,Wiseman:2011by} 
in which case, instead of (\ref{Einstein-eqs}), 
one solves the so called EDT equations
\begin{eqnarray}
\label{EDT}
R_{\mu\nu}-\nabla_{(\mu}\xi_{\nu)}=-\frac{3}{L^2} g_{\mu\nu}+ 
2\left(T_{\mu\nu}-\frac{1}{2}T  g_{\mu\nu}\right) \ .
\end{eqnarray}
Here, $\xi^\mu$ is a vector defined as 
$
\xi^\mu\equiv g^{\nu\rho}(\Gamma_{\nu\rho}^\mu-\bar \Gamma_{\nu\rho}^\mu)\ ,
$
where 
$\Gamma_{\nu\rho}^\mu$ is the Levi-Civita connection associated to the
spacetime metric $g$ that one wants to determine, and a reference metric $\bar g$ is introduced, 
$\bar \Gamma_{\nu\rho}^\mu$ being the corresponding Levi-Civita connection.
Solutions to (\ref{EDT}) also solve the Einstein equations
iff $\xi^\mu \equiv 0$ everywhere.
To achieve this,
we impose boundary conditions  which are compatible with
$\xi^\mu = 0$
on the boundary of the domain of integration.
Then, this should imply $\xi^\mu \equiv 0$ everywhere,
a condition which is verified from  the numerical output.

In our approach, 
%we have found convenient to 
%%work with horospherical coordinates
we use  a metric ansatz
with five unknown functions,
 $\{ {\cal F}_0,{\cal F}_1,{\cal F}_2,{\cal F}_3,S_1  \}$,
\begin{eqnarray}
\label{metric} 
&&
ds^2=-{\cal F}_0(x,\theta)\frac{T_0(x)^2}{T_1(x)^2}dt^2+{\cal F}_1(x,\theta)\frac{dx^2}{T_1(x)^2}
+{\cal F}_2(x,\theta)\frac{x^2}{T_1(x)^2}\left[ d\theta+\frac{S_1(x,\theta)}{x}dx\right]^2 
\nonumber 
\\
&&
{~~~~~~~}
+{\cal F}_3(x,\theta)\frac{x^2\sin^2\theta}{T_1(x)^2}  d\varphi^2 \ ,~~~~~
{\rm where}~~~~T_0(x)=\frac{1+x^2}{2L},~~T_1(x)=\frac{1-x^2}{2L}~,
\end{eqnarray}
with $0\leq x \leq 1$.
Setting 
$ {\cal F}_0={\cal F}_1={\cal F}_2={\cal F}_3=1,$ $S_1=0$
results in global $AdS$ metric written in horospherical coordinates, which can be obtained by taking
 $r=2L\frac{x}{1-x^2}$ in the line element (\ref{AdS}), 
providing the obvious reference metric $\bar g$.

The matter Ansatz is still given by (\ref{gauge-ansatz}),
with a single (magnetic) potential  $\Phi(x,\theta)$.
As such,
the EDT equations (\ref{EDT}) together with vector field equations (\ref{F-eqs})
result in a set of six elliptic partial differential equations which are solved numerically
as a boundary value problem.
The boundary conditions are found by constructing an approximate form
of the solutions on the boundary of the domain
of integration compatible with regularity of the solutions and the requirement $\xi^\mu = 0$.

In what follows we shall present  results for 
simplest $k=1$, $n=0$ case, associated with the (probe limit) 
first branch in Figure \ref{profile}.
There the boundary conditions we impose are
\begin{eqnarray}
\label{bcP}
\Phi|_{x=0}=0,~\partial_\theta \Phi \big|_{\theta=0,\pi}=0~,
~
\Phi \big|_{x=1}=c_0 \sin^2 \theta,
\end{eqnarray} 
(with $c_0>0$ an input parameter)
for the magnetic potential,
and
\begin{eqnarray}
\nonumber 
 \partial_x {\cal F}_i\big|_{x=0}
%=\partial_x {\cal F}_1\big|_{x=0}
%=\partial_x {\cal F}_2\big|_{x=0}
%=\partial_x {\cal F}_3\big|_{x=0}
=\partial_x S_1\big|_{x=0} 
=0, 
~~~
 \partial_\theta {\cal F}_i\big|_{\theta=0,\pi}
%=\partial_\theta {\cal F}_1\big|_{\theta=0,\pi}
%=\partial_\theta {\cal F}_2\big|_{\theta=0,\pi}=
%=\partial_\theta {\cal F}_3\big|_{\theta=0,\pi}=
= S_1\big|_{\theta=0,\pi}
=0,
~~~
  {\cal F}_i\big|_{x=1} =1,~~
%= {\cal F}_1\big|_{x=1}
%= {\cal F}_2\big|_{x=1} 
%= {\cal F}_3\big|_{x=1} =1,~~ 
S_1\big|_{x=1}=0,
\end{eqnarray} 
for the metric functions, with $i=0,1,2,3$.
Moreover, we shall assume  that the solutions are
symmetric $w.r.t.$ a reflection in the
equatorial plane, which implies
that 
the functions ${\cal F}_i$ and $\Phi$
satisfy Neumann boundary conditions at $\theta=\pi/2$,
while $S_1$  
vanishes there.  

While the Noether charge is still given by 
(\ref{q}),
the computation of the mass $M$ of the solutions 
is more complicated.
In what follows,
$M$
is computed by employing the boundary counterterm 
approach in \cite{Balasubramanian:1999re}, 
being the conserved charges associated with Killing symmetry
$\partial_t$ of the induced boundary metric.
%  found for $x=$constant (and close to one).
%
This results in
 the following expression of
the  mass:
\begin{eqnarray}
\nonumber
M=-\frac{3 L }{ 8 G }
\int_0^\pi d\theta   \sin \theta  (f_{23}(\theta)+f_{33}(\theta)) ,
\end{eqnarray} 
where 
$f_{23}(\theta)$,
$f_{33}(\theta)$
are two functions which enter  the asymptotic form of the
metric functions,
${\cal F}_2=1+ f_{23}(\theta)(1-x)^3+\dots $,
${\cal F}_3=1+ f_{33}(\theta)(1-x)^3+\dots $
(also
${\cal F}_0=1- (f_{23}(\theta)+f_{33}(\theta))(1-x)^3+\dots $),
being found from the numerical output.

%\bigskip
%%%%%%%%%%%%%%%%%%%%%%%%%%%%%%%%%%%%%%%%%%%%%%%%%%%%%%%%%%%%%%%%%%%%%%%%%%%%%%
 \begin{figure}[h!]
\begin{center}
\includegraphics[width=0.45\textwidth]{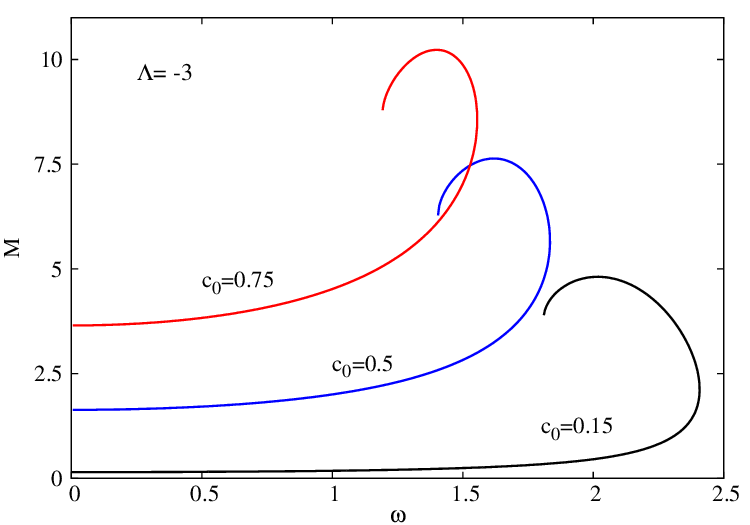}
\includegraphics[width=0.45\textwidth]{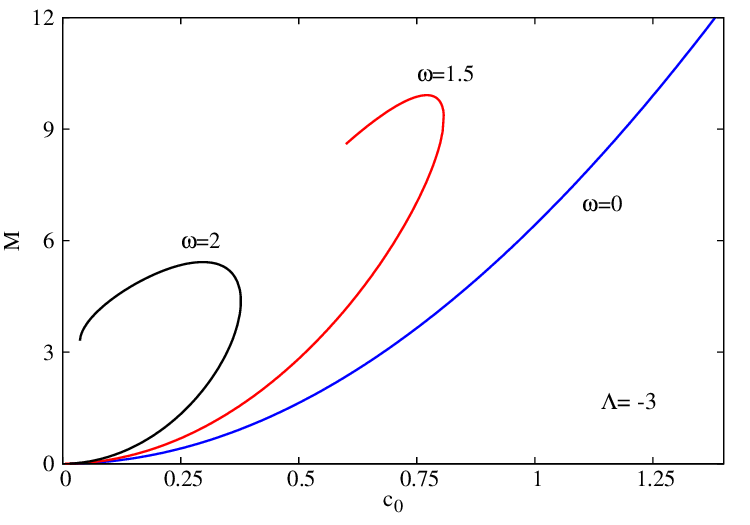}
\includegraphics[width=0.45\textwidth]{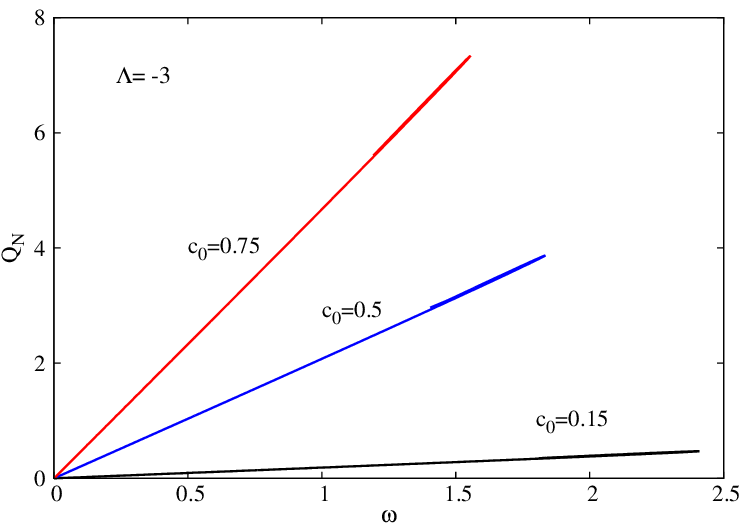}
\includegraphics[width=0.45\textwidth]{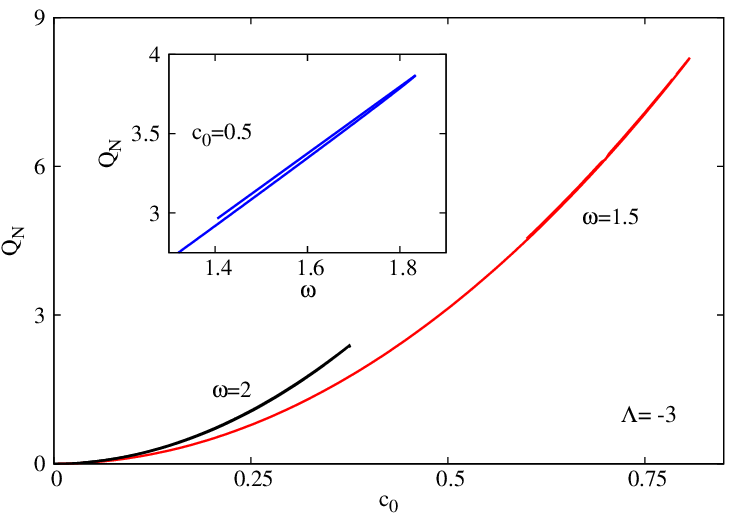}
\caption{ 
The mass and the Noether charge of the  
gravitating generalization of the
 scalar clouds with $k=1$, $n=0$ 
 are plotted as a function of frequency 
$\omega$
(left panel) and as a function of the (maximal) magnitude of the 
magnetic potential at infinity $c_0$
(right panel).  
}
\label{fig2}
\end{center}
\end{figure} 
%%%%%%%%%%%%%%%%%%%%%%%%%%%%%%%%%%%%%%%%%%%%%%%%%%%%%%%%%%%%%%%%%%%%%%%%%%%%%%

In this approach, the only input parameters are 
the field frequency $\omega$,
the asymptotic magnitude $c_0$ of the vector potential in the equatorial plane
and
the cosmological length scale 
$L$, which we take $L=1$.
The  equations are discretized in an equidistant ($x,\theta$)-grid, 
with (usually) around $250\times 50$ points.
Using the  Newton-Raphson approach, the resulting system is solved iteratively 
until convergence is achieved~\footnote{We have employed a finite difference method, 
with  a sixth order for the difference formulae.}.
Also, the typical  numerical error
  is estimated to be of the order of $10^{-5}$,
	except for  the secondary branches of solutions 
	(see below).

 Fixing the  boundary data at $x=1$,
$i.e.$ the parameter $c_0$ in  (\ref{bcP}), 
and increasing the value of frequency,
a branch of solutions smoothly emerges from 
any
static Einstein-Maxwell  soliton. 
Along this branch, both mass and Noether charge increase with $\omega$,
see Figure \ref{fig2} (left panel). 
These solutions stop 
to exist for 
some critical frequency,
which decreases with increasing $c_0$
(being always smaller than $\omega_{c}^{(1)}$ as defined by (\ref{wcrit}).
 There a new branch of solutions emerges,
extending backwards in $\omega$, the maximal value of
mass being approached along this branch.
For any $c_0$,
this (secondary) branch stops to exist
for another critical value of the frequency.
We mention that the numerical accuracy decreases significantly along this branch;
however, as with the usual boson stars  \cite{Liebling:2012fv},  
 we expect the occurrence of (other)
secondary branches of solutions, with an inspiraling behaviour
of the curve $M(\omega)$.

Interestingly, the picture obtained for 
fixed $\omega$ and increasing $c_0$
resembles the situation above.
Again, the solutions with  $\omega>0$ stop to exist for 
a maximal $c_0$,
where a secondary branch of solutions
extending backwards in $c_0$ emerges.

%%%%%%%%%%%%%%%%%%%%%%%%%%%%%%%%%%%%%%%%%%%%%%%%%%%%%%%%%%%%%%%%%%%%%%%%%%%%%
\section{Further remarks. Conclusions}
\label{sec_5}
%%%%%%%%%%%%%%%%%%%%%%%%%%%%%%%%%%%%%%%%%%%%%%%%%%%%%%%%%%%%%%%%%%%%%%%%%%%%sks}
 
The main purpose of this work was to show that
the $D=4$  
 Einstein gravity with negative cosmological constant
 minimally coupled to complex Maxwell field  allows for boson-star--like solitonic solutions, which are static and axially symmetric.
This is a consequence of endowing the vector potentials
with a specific harmonic time dependence, which, however, 
is not present in the total energy-momentum tensor.
The considered model  corresponds to the massless limit of the Einstein-complex-Proca-AdS theory.
Given some appropriate boundary data at infinity,  
the EcM
  solitonic solutions depend on a single input parameter,
	which is the field frequency,
	with two global charges, the mass and the total number of particles (the Noether charge).
Their existence can be traced back to the ``box"-like behaviour of the AdS spacetime\footnote{
  {It is interesting to contrast the situation with that found for a model with a massless, non-self-interacting, complex scalar field ($i.e.$ two real scalars).
  Again, the scalar clouds on a fixed $AdS$ background can be found in closed form, for any multipole moment.
  However, in contrast to the vector case, they possess finite charges for
  a discrete set of frequencies
  $\omega\sim 1/L>0$, only,
  without a static limit.
The nonlinear continuation of these solutions corresponds to $AdS$ (scalar) boson stars
(note that only spherically symmetric solutions have been studied so far \cite{Buchel:2013uba}).
  }
  }.
	 No analogue configurations exist for an asymptotically flat spacetime

The work here has restricted to the simplest ansatz
with a single magnetic potential\footnote{ We mention that,
as with the Maxwell case,
one can consider a dual description of the solutions in this work.
The corresponding ansatz is more complicated, 
 with both electric and magnetic potentials,
$ {\cal A}= ( iV(r,\theta) dt +H_1(r,\theta) dr +H_2(r,\theta) d\theta)e^{-i \omega t} $,
where
$V,H_1$ and $H_2$ are real functions determined by  $\Phi$.
}.
However, we expect the same model to possess a variety of other more complicated solutions.
In particular, following the Refs. 
\cite{Herdeiro:2016plq,Herdeiro:2020kvf},
we predict the existence of boson-star--like  configurations
that have no continuous spatial symmetries. 
 
\medskip

A question which arises naturally is if one can add a small black hole (BH)
at the center of the solitons considered in this work.
This is indeed the case for the $\omega=0$ solutions,
where in addition to the (magnetic) Reissner-Nordstr\"om-AdS BHs,
one also finds static BHs solutions with an arbitrary multipole structure, 
including configurations that have no continuous spatial symmetries\footnote{This corresponds  
to considering the magnetic duals of the BHs with an electric potential
  in Ref. \cite{Herdeiro:2016plq}.}. 
Although we do not have a rigorous proof,  
the situation seems to be different for
configurations 
with $\omega \neq 0$.
Some hints in this direction  follow from a study of the eqs. (\ref{F-eqs})
in a fixed static BH background.
There one assumes the existence of a power series expansion for $\Phi(r,\theta)$
 in the vicinity the event horizon, $\Phi(r,\theta) =\sum_{p \geq 0}\phi_p(\theta) (r-r_H)^p$ 
(with $r=r_H >0$ the event horizon radius).
Then, when replacing in eq. (\ref{F-eqs}), it follows that 
the coefficients $\phi_p$ vanish order by order. 
This is essentially the picture found for configurations
with (massive, complex) 
	scalar
\cite{Pena:1997cy,Astefanesei:2003qy,Yazadjiev:2024rql},
	or vector 
	\cite{Herdeiro:2016tmi}
 fields endowed with a  harmonic time dependence and a
 static geometry, and a rigorous non-existence  
	proof could be found in those cases.
 However, we predict the existence of 
BHs solutions with $\omega\neq 0$ for a more general framework
 allowing for rotation \cite{Herdeiro:2016tmi},
or by generalizing  the action (\ref{actionN})
for a model with a $U(1)$-gauged
 complex vector field  
and
 extending the construction in 
\cite{SalazarLandea:2016bys,Herdeiro:2020xmb}.

\medskip
 Returning to the general EcM case,
it would be interesting to find 
  a generalization of the action 
(\ref{action})
within a gauged supergravity model.
A truncation of the 11-dimensional supergravity
that leads to the $D=4$ Einstein-$U(1)^2$  action with negative cosmological constant 
is discussed in Ref. \cite{Marolf:2021kjc}.
However, the two Maxwell fields there are
subject to a constraint which is not satisfied by the solutions in this work.
A different setting is provided by the 
$N=4$ gauged $SU(2) \times SU(2)$ supergravity
\cite{Freedman:1978ra}
which contains two (non-Abelian) gauge fields  
with
 the same coupling with a dilaton and an axion.
However, in this case the dilaton field possesses a Liouville potential
which does not allow for an $AdS$ vacuum.

%Clarify the physical interpretation of the Minkowski spacetime configurations.

%%%%%%%%%%%%%%%%%%%%%%%%%%%%%%%%%%%%%%%%%%%%%%%%%%%%%%%%%%%%%%%%%%%%%%%%%%%%%%%%%%%%%%%%%%%%%%%%%%%%%%
\section*{Acknowledgements}
%%%%%%%%%%%%%%%%%%%%%%%%%%%%%%%%%%%%%%%%%%%%%%%%%%%%%%%%%%%%%%%%%%%%%%%%%%%%%%%%%%%%%%%%%%%%%%%%%%%%%%
 
The work of C.R. and E.R.
is supported by the Center for Research and Development in Mathematics and Applications (CIDMA) 
through the Portuguese Foundation for Science and Technology (FCT -- Fundaç\~ao para a Ci\^encia e a Tecnologia)
 through projects: UIDB/04106/2020 (DOI identifier \url{https://doi.org/10.54499/UIDB/04106/2020}); UIDP/04106/2020 (DOI identifier \url{https://doi.org/10.54499/UIDP/04106/2020});  PTDC/FIS-AST/3041/2020 (DOI identifier \url{http://doi.org/10.54499/PTDC/FIS-AST/3041/2020}); CERN/FIS-PAR/0024/2021 (DOI identifier \url{http://doi.org/10.54499/CERN/FIS-PAR/0024/2021}); and 2022.04560.PTDC (DOI identifier \url{https://doi.org/10.54499/2022.04560.PTDC}).   
 This work has further been supported by the European Horizon Europe staff exchange (SE) programme HORIZON-MSCA2021-SE-01 Grant No. NewFunFiCO101086251.
We also garatefully acknowledge support by DFG project Ku612/18-1 and the National Natural
Science Foundation of China (NSFC) Grant No. 12205123 and Jiangxi Provincial Natu-
ral Science Foundation with Grant No. 20232BAB211029 and by the Sino-German (CSC-
DAAD) Postdoc Scholarship Program, No. 2021 (57575640)

%%%%%%%%%%%%%%%%%%%%%%%%%%%%%%%%%%%%%%%%%%%%%%%%%%%%%%%%%%%%%%%%%%%%%%%%%%%%%%%%%%%%%  
 \begin{small}
 
%%%%%%%%%%%%%%%%%%%%%%%%%%%%%%%%%%%%%%%%%%%%%%%%%%%%%%%%%%%%%%%%%%%%%%%%%%%%%%
 \end{small}


\begin{thebibliography}{99}
%%%%%%%%%%%%%%%%%%%%%%%%%%%%%%%%%%%%%%%%%%%%%%%%%%%%%%%%%%%%%%%%%%%%%%%%%%%%%%%%%%%%%


%\cite{Maldacena:1997re}
\bibitem{Maldacena:1997re}
  J.~M.~Maldacena,
  %``The Large N limit of superconformal field theories and supergravity,''
  Int.\ J.\ Theor.\ Phys.\  {\bf 38} (1999) 1113
   [Adv.\ Theor.\ Math.\ Phys.\  {\bf 2} (1998) 231]
  [hep-th/9711200].
  %%CITATION = HEP-TH/9711200;%%
  %10691 citations counted in INSPIRE as of 24 Apr 2015
%%%%%%%%%%%%%%%%%%%%%%%%%%%%%%%%%%%%%%%%%%%%%%%%%%%%%%%%%%%%%%%%%%%%%%%%%%%%%%%%%%%%%    

%\cite{Ruffini:1971bza}
\bibitem{Ruffini:1971bza}
R.~Ruffini and J.~A.~Wheeler,
%``Introducing the black hole,''
Phys. Today \textbf{24} (1971) no.1, 30
%doi:10.1063/1.3022513
%509 citations counted in INSPIRE as of 04 May 2024

%\cite{Herdeiro:2015vaa} 
\bibitem{Herdeiro:2015vaa}
  C.~Herdeiro and E.~Radu,
  %``Anti-de-Sitter regular electric multipoles: Towards Einstein–Maxwell-AdS solitons,''
  Phys.\ Lett.\ B {\bf 749} (2015) 393
%  doi:10.1016/j.physletb.2015.08.010
  [arXiv:1507.04370 [gr-qc]].
  %%CITATION = doi:10.1016/j.physletb.2015.08.010;%%
  %19 citations counted in INSPIRE as of 31 Oct 2019
%%%%%%%%%%%%%%%%%%%%%%%%%%%%%%%%%%%%%%%%%%%%%%%%%%%%%%%%%%%%%%%%%%%%%%%%%%%% 
%\cite{Costa:2015gol}
\bibitem{Costa:2015gol}
M.~S.~Costa, L.~Greenspan, M.~Oliveira, J.~Penedones and J.~E.~Santos,
%``Polarised Black Holes in AdS,''
Class. Quant. Grav. \textbf{33} (2016) no.11, 115011
%doi:10.1088/0264-9381/33/11/115011
[arXiv:1511.08505 [hep-th]].
%24 citations counted in INSPIRE as of 23 Apr 2024
%%%%%%%%%%%%%%%%%%%%%%%%%%%%%%%%%%%%%%%%%%%%%%%%%%%%%%%%%%%%%%%%%%%%%%%%%%%% 
%\cite{Herdeiro:2016xnp}
\bibitem{Herdeiro:2016xnp}
  C.~Herdeiro and E.~Radu,
  %``Einstein–Maxwell–Anti-de-Sitter spinning solitons,''
  Phys.\ Lett.\ B {\bf 757} (2016) 268
 % doi:10.1016/j.physletb.2016.04.004
  [arXiv:1602.06990 [gr-qc]].
  %%CITATION = doi:10.1016/j.physletb.2016.04.004;%%
  %11 citations counted in INSPIRE as of 02 Nov 2019
%%%%%%%%%%%%%%%%%%%%%%%%%%%%%%%%%%%%%%%%%%%%%%%%%%%%%%%%%%%%%%%%%%%%%%%%%%%% 
%\cite{Herdeiro:2016plq}
\bibitem{Herdeiro:2016plq}
  C.~A.~R.~Herdeiro and E.~Radu,
  %``Static Einstein-Maxwell black holes with no spatial isometries in AdS space,''
  Phys.\ Rev.\ Lett.\  {\bf 117} (2016) no.22,  221102
 % doi:10.1103/PhysRevLett.117.221102
  [arXiv:1606.02302 [gr-qc]].
  %%CITATION = doi:10.1103/PhysRevLett.117.221102;%%
  %13 citations counted in INSPIRE as of 02 Nov 2019 
%%%%%%%%%%%%%%%%%%%%%%%%%%%%%%%%%%%%%%%%%%%%%%%%%%%%%%%%%%%%%%%%%%%%%%%%%%%% 
%\cite{Liebling:2012fv}
\bibitem{Liebling:2012fv}
S.~L.~Liebling and C.~Palenzuela,
%``Dynamical boson stars,''
Living Rev. Rel. \textbf{26} (2023) no.1, 1
%doi:10.1007/s41114-023-00043-4
[arXiv:1202.5809 [gr-qc]].
%629 citations counted in INSPIRE as of 29 Apr 2024
%%%%%%%%%%%%%%%%%%%%%%%%%%%%%%%%%%%%%%%%%%%%%%%%%%%%%%%%%%%%%%%%%%%%%%%%%%%% 
%\cite{Duarte:2016lig}
\bibitem{Duarte:2016lig}
M.~Duarte and R.~Brito,
%``Asymptotically anti-de Sitter Proca Stars,''
Phys. Rev. D \textbf{94} (2016) no.6, 064055
%doi:10.1103/PhysRevD.94.064055
[arXiv:1609.01735 [gr-qc]].
%25 citations counted in INSPIRE as of 08 Apr 2024
%%%%%%%%%%%%%%%%%%%%%%%%%%%%%%%%%%%%%%%%%%%%%%%%%%%%%%%%%%%%%%%%%%%%%%%%%%%%
%\cite{Astefanesei:2003qy}  
\bibitem{Astefanesei:2003qy}
D.~Astefanesei and E.~Radu,
%``Boson stars with negative cosmological constant,''
Nucl. Phys. B \textbf{665} (2003), 594-622
%doi:10.1016/S0550-3213(03)00482-6
[arXiv:gr-qc/0309131 [gr-qc]].
%107 citations counted in INSPIRE as of 29 Apr 2024
%%%%%%%%%%%%%%%%%%%%%%%%%%%%%%%%%%%%%%%%%%%%%%%%%%%%%%%%%%%%%%%%%%%%%%%%%%%%
%\cite{Buchel:2013uba}
\bibitem{Buchel:2013uba}
A.~Buchel, S.~L.~Liebling and L.~Lehner,
%``Boson stars in AdS spacetime,''
Phys. Rev. D \textbf{87} (2013) no.12, 123006
%doi:10.1103/PhysRevD.87.123006
[arXiv:1304.4166 [gr-qc]].
%164 citations counted in INSPIRE as of 29 Apr 2024
%%%%%%%%%%%%%%%%%%%%%%%%%%%%%%%%%%%%%%%%%%%%%%%%%%%%%%%%%%%%%%%%%%%%%%%%%%%%  
%\cite{Brito:2015pxa}
\bibitem{Brito:2015pxa}
R.~Brito, V.~Cardoso, C.~A.~R.~Herdeiro and E.~Radu,
%``Proca stars: Gravitating Bose\textendash{}Einstein condensates of massive spin 1 particles,''
Phys. Lett. B \textbf{752} (2016), 291-295
%doi:10.1016/j.physletb.2015.11.051
[arXiv:1508.05395 [gr-qc]].
%241 citations counted in INSPIRE as of 29 Apr 2024
%%%%%%%%%%%%%%%%%%%%%%%%%%%%%%%%%%%%%%%%%%%%%%%%%%%%%%%%%%%%%%%%%%%%%%%%%%%%  
%\cite{Headrick:2009pv}
\bibitem{Headrick:2009pv}
  M.~Headrick, S.~Kitchen and T.~Wiseman,
  %``A New approach to static numerical relativity, and its application to Kaluza-Klein black holes,''
  Class.\ Quant.\ Grav.\  {\bf 27} (2010) 035002
  [arXiv:0905.1822 [gr-qc]].
  %%CITATION = ARXIV:0905.1822;%%
  %46 citations counted in INSPIRE as of 04 Apr 2014
%%%%%%%%%%%%%%%%%%%%%%%%%%%%%%%%%%%%%%%%%%%%%%%%%%%%%%%%%%%%%%%%%%%%%%%%%%%%   
%\cite{Adam:2011dn}
\bibitem{Adam:2011dn}
  A.~Adam, S.~Kitchen and T.~Wiseman,
  %``A numerical approach to finding general stationary vacuum black holes,''
  Class.\ Quant.\ Grav.\  {\bf 29} (2012) 165002
  [arXiv:1105.6347 [gr-qc]].
  %%CITATION = ARXIV:1105.6347;%%
  %8 citations counted in INSPIRE as of 04 Apr 2014
 %%%%%%%%%%%%%%%%%%%%%%%%%%%%%%%%%%%%%%%%%%%%%%%%%%%%%%%%%%%%%%%%%%%%%%%%%%%%
 %\cite{Wiseman:2011by}
\bibitem{Wiseman:2011by}
  T.~Wiseman,
  %``Numerical construction of static and stationary black holes,''
  arXiv:1107.5513 [gr-qc].
  %%CITATION = ARXIV:1107.5513;%%
  %11 citations counted in INSPIRE as of 04 Apr 2014  
%%%%%%%%%%%%%%%%%%%%%%%%%%%%%%%%%%%%%%%%%%%%%%%%%%%%%%%%%%%%%%%%%%%%%%%%%%%%       
 %\cite{Balasubramanian:1999re}
\bibitem{Balasubramanian:1999re}
  V.~Balasubramanian and P.~Kraus,
  %``A Stress tensor for Anti-de Sitter gravity,''
  Commun.\ Math.\ Phys.\  {\bf 208} (1999) 413
  [hep-th/9902121].
  %%CITATION = HEP-TH/9902121;%%
  %1006 citations counted in INSPIRE as of 24 Apr 2015 	
	 %%%%%%%%%%%%%%%%%%%%%%%%%%%%%%%%%%%%%%%%%%%%%%%%%%%%%%%%%%%%%%%%%%%%%%%%%%%%       
%\cite{Herdeiro:2020kvf}
\bibitem{Herdeiro:2020kvf}
C.~A.~R.~Herdeiro, J.~Kunz, I.~Perapechka, E.~Radu and Y.~Shnir,
%``Multipolar boson stars: macroscopic Bose-Einstein condensates akin to hydrogen orbitals,''
Phys. Lett. B \textbf{812} (2021), 136027
%doi:10.1016/j.physletb.2020.136027
[arXiv:2008.10608 [gr-qc]].
%40 citations counted in INSPIRE as of 02 May 2024 
 %%%%%%%%%%%%%%%%%%%%%%%%%%%%%%%%%%%%%%%%%%%%%%%%%%%%%%%%%%%%%%%%%%%%%%%%%%%%	
%\cite{Pena:1997cy}
\bibitem{Pena:1997cy}
I.~Pena and D.~Sudarsky,
%``Do collapsed boson stars result in new types of black holes?,''
Class. Quant. Grav. \textbf{14} (1997), 3131-3134
%doi:10.1088/0264-9381/14/11/013
%83 citations counted in INSPIRE as of 29 Apr 2024
%%%%%%%%%%%%%%%%%%%%%%%%%%%%%%%%%%%%%%%%%%%%%%%%%%%%%%%%%%%%%%%%%%%%%%%%%%%%  
%\cite{Yazadjiev:2024rql}
\bibitem{Yazadjiev:2024rql}
S.~Yazadjiev and D.~Doneva,
%``Black hole no-hair theorem for self-gravitating time-dependent spherically symmetric multiple scalar fields,''
[arXiv:2401.13288 [gr-qc]].
%0 citations counted in INSPIRE as of 29 Apr 2024
%%%%%%%%%%%%%%%%%%%%%%%%%%%%%%%%%%%%%%%%%%%%%%%%%%%%%%%%%%%%%%%%%%%%%%%%%%%%  
%\cite{Herdeiro:2016tmi}
\bibitem{Herdeiro:2016tmi}
C.~Herdeiro, E.~Radu and H.~R\'unarsson,
%``Kerr black holes with Proca hair,''
Class. Quant. Grav. \textbf{33} (2016) no.15, 154001
%doi:10.1088/0264-9381/33/15/154001
[arXiv:1603.02687 [gr-qc]].
%240 citations counted in INSPIRE as of 29 Apr 2024
%%%%%%%%%%%%%%%%%%%%%%%%%%%%%%%%%%%%%%%%%%%%%%%%%%%%%%%%%%%%%%%%%%%%%%%%%%%%
%\cite{SalazarLandea:2016bys}
\bibitem{SalazarLandea:2016bys}
I.~Salazar Landea and F.~Garc\'\i{}a,
%``Charged Proca Stars,''
Phys. Rev. D \textbf{94} (2016) no.10, 104006
%doi:10.1103/PhysRevD.94.104006
[arXiv:1608.00011 [hep-th]].
%42 citations counted in INSPIRE as of 03 May 2024
 %%%%%%%%%%%%%%%%%%%%%%%%%%%%%%%%%%%%%%%%%%%%%%%%%%%%%%%%%%%%%%%%%%%%%%%%%%%%
%\cite{Herdeiro:2020xmb}
\bibitem{Herdeiro:2020xmb}
C.~A.~R.~Herdeiro and E.~Radu,
%``Spherical electro-vacuum black holes with resonant, scalar $Q$-hair,''
Eur. Phys. J. C \textbf{80} (2020) no.5, 390
%doi:10.1140/epjc/s10052-020-7976-9
[arXiv:2004.00336 [gr-qc]].
%30 citations counted in INSPIRE as of 03 May 2024



%%%%%%%%%%%%%%%%%%%%%%%%%%%%%%%%%%%%%%%%%%%%%%%%%%%%%%%%%%%%%%%%%%%%%%%%%%%% 
%\cite{Marolf:2021kjc}
\bibitem{Marolf:2021kjc}
D.~Marolf and J.~E.~Santos,
%``AdS Euclidean wormholes,''
Class. Quant. Grav. \textbf{38} (2021) no.22, 224002
%doi:10.1088/1361-6382/ac2cb7
[arXiv:2101.08875 [hep-th]].
%51 citations counted in INSPIRE as of 21 Apr 2024 
%%%%%%%%%%%%%%%%%%%%%%%%%%%%%%%%%%%%%%%%%%%%%%%%%%%%%%%%%%%%%%%%%%%%%%%%%%%% 
%\cite{Freedman:1978ra}
\bibitem{Freedman:1978ra}
D.~Z.~Freedman and J.~H.~Schwarz,
%``N=4 Supergravity Theory with Local SU(2) x SU(2) Invariance,''
Nucl. Phys. B \textbf{137} (1978), 333-339
%doi:10.1016/0550-3213(78)90526-6
%183 citations counted in INSPIRE as of 29 Apr 2024
      

     
 \end{thebibliography}
\end{document}